\documentclass[english,aps,twocolumn,superscriptaddress,apl]{revtex4-1}
\usepackage[T1]{fontenc}
\usepackage[latin9]{inputenc}
\usepackage{units}
\usepackage{textcomp}
\usepackage{graphicx}

\begin{document}

\title{Bunches of misfit dislocations on the onset of relaxation of Si$_{0.4}$Ge$_{0.6}$/Si(001)
epitaxial films revealed by high-resolution x-ray diffraction}

\author{Vladimir Kaganer}

\address{Paul-Drude-Institut für Festkörperelektronik, Hausvogteiplatz 5--7, 10117 Berlin, Germany}

\author{Tatjana Ulyanenkova}

\address{Rigaku Europe SE, Am Hardwald 11, 76275 Ettlingen, Germany}

\author{Andrei Benediktovitch}

\address{Atomicus OOO, Mogilevskaya Str.\ 39a-530, 220007 Minsk, Belarus}

\author{Maksym Myronov}

\address{The University of Warwick, Department of Physics, Coventry CV4 7AL, UK}

\author{Alex Ulyanenkov}

\address{Atomicus GmbH, Schoemperlen Str.\ 12a, 76185 Karlsruhe, Germany}

\begin{abstract}
The experimental x-ray diffraction patterns of a Si$_{0.4}$Ge$_{0.6}$/Si(001)
epitaxial film with a low density of misfit dislocations are modeled
by the Monte Carlo method. It is shown that an inhomogeneous distribution
of 60$^\circ$ dislocations with dislocations arranged in bunches
is needed to explain the experiment correctly. As a result of the
dislocation bunching, the positions of the x-ray diffraction peaks
do not correspond to the average dislocation density but reveal less
than a half of the actual relaxation.
\end{abstract}

\maketitle

Si$_{1-x}$Ge$_{x}$ films on Si substrate constitute a heteroepitaxial
system that finds numerous applications in the whole compositional range
\cite{cressler06} and, at the same time, is
a model system that demonstrates the whole spectrum of strain relaxation
mechanisms. When either thickness or Ge content is small, the layers
stay strained, by accepting the lateral lattice spacing of the
substrate and expanding vertically due to the Poisson effect. Larger
strain is relaxed by one of the two mechanisms, plastic relaxation
in planar layers by introduction of misfit dislocations at the interface
\cite{fitzgerald91,hull92,jain97,bolkhovityanov01} or development
of three-dimensional islands in the Stranski\textendash Krastanov
growth mode \cite{brunner02}. Planar films with controlled strain
state are required for various applications. Particularly, the high
compressive strain in the Si$_{0.4}$Ge$_{0.6}$ films, studied in
the present work, ought to enhance a room-temperature two-dimensional
hole gas mobility, which is important for application of such films
in the field effect transistors\cite{myronov14}.

X-ray diffraction is a well established technique to characterize
relaxed epitaxial films. Positions of the x-ray peaks provide the
lattice parameters of a relaxed film and hence the density of misfit
dislocations \cite{heinke94,bowen06,benediktovich14}. An application
of the same analysis at the onset of relaxation suffers from the peak
broadening due to a small layer thickness \cite{hartmann11}. Moreover,
we show below that the inhomogeneity in the dislocation distribution
plays an essential role in the x-ray diffraction analysis. The position
of the coherent peak is given by the less strained regions of the
film and hence underestimates the relaxation. The diffuse x-ray intensity
occurs more sensitive to both the presence of misfit dislocations
and their distribution.

We have chosen for a detailed x-ray diffraction study a 27~nm thick
Si$_{0.4}$Ge$_{0.6}$ film on Si(001), demonstrating an early stage
of the relaxation. Thinner unrelaxed and thicker relaxed films of
the same series of samples were studied earlier \cite{ulyanenkova13}.
The samples were grown by reduced pressure chemical vapor deposition
in an industrial ASM Epsilon 2000 system. Germane and disilane precursors
were used to grow Si$_{0.4}$Ge$_{0.6}$ epilayers at the growth temperature
of 450$^{\circ}$\,C. The critical thickness for plastic relaxation
at a Ge content $x=0.6$ is 10~nm \cite{people85,people86,hartmann11}.
The low growth temperature allows us to obtain 2.7 times thicker
layer possessing a small relaxation.

High-resolution x-ray diffraction measurements were performed
using a 9~kV SmartLab Rigaku diffractometer with a rotating anode.
The diffraction setup included a two-crystal Ge monochromator in the
400 setting and a one-dimensional high-speed position-sensitive detector
D/teX Ultra from Rigaku.

\begin{figure*}
\includegraphics[width=1\textwidth]{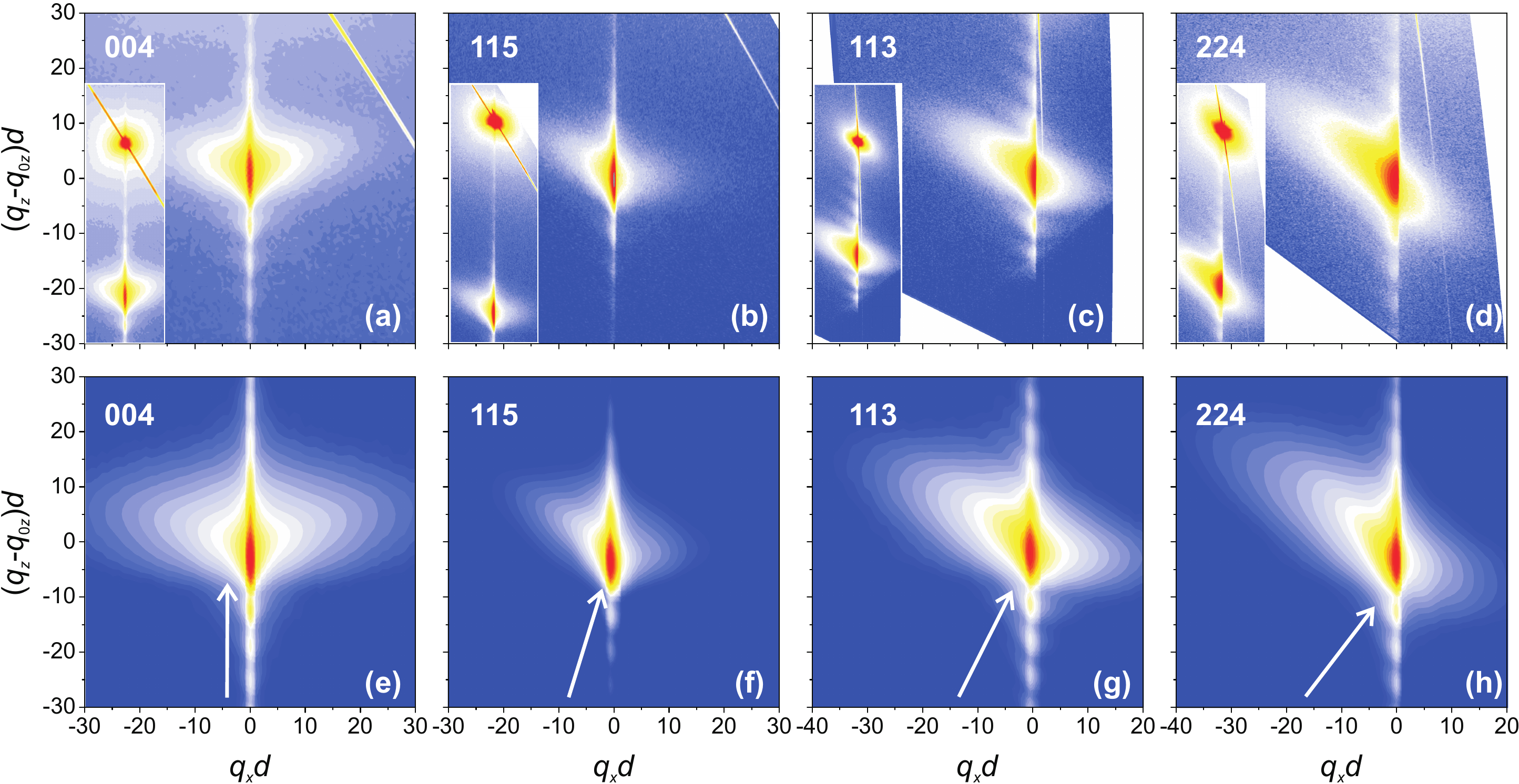}
\caption{Experimental (a-d) and Monte Carlo simulated (e-h) reciprocal space
maps of a 27~nm thick Si$_{0.4}$Ge$_{0.6}$ film on Si(001). The
directions of the scattering vectors are shown by white arrows. The
insets in (a-d) show full reciprocal space maps comprising the substrate
and the film peaks.}
\label{fig:maps}
\end{figure*}

Figures \ref{fig:maps}(a-d) present experimental reciprocal space
maps in the symmetric 004 and several asymmetric reflections, in a
sequence of increasing asymmetry. The wave vectors are represented
in the dimensionless units of the product of the components of the
reciprocal space vector $(q_{x},q_{z})$ and the film thickness $d=27$~nm.
Each map comprises a coherent scattering streak extended along the
surface normal, and diffuse scattering. The presence of the coherent
and the diffuse intensities is an indication of a weakly distorted
film, possessing a low density of misfit dislocations. A closer inspection
of the symmetric 004 map in Fig.~\ref{fig:maps}(a) reveals that
the positions of the coherent and the diffuse maxima do not coincide:
with the origin $q_{z}=0$ chosen at the position of the coherent
peak, the diffuse intensity is maximum at $q_{z}d\approx5.5$. 

\begin{figure}
\includegraphics[width=0.9\columnwidth]{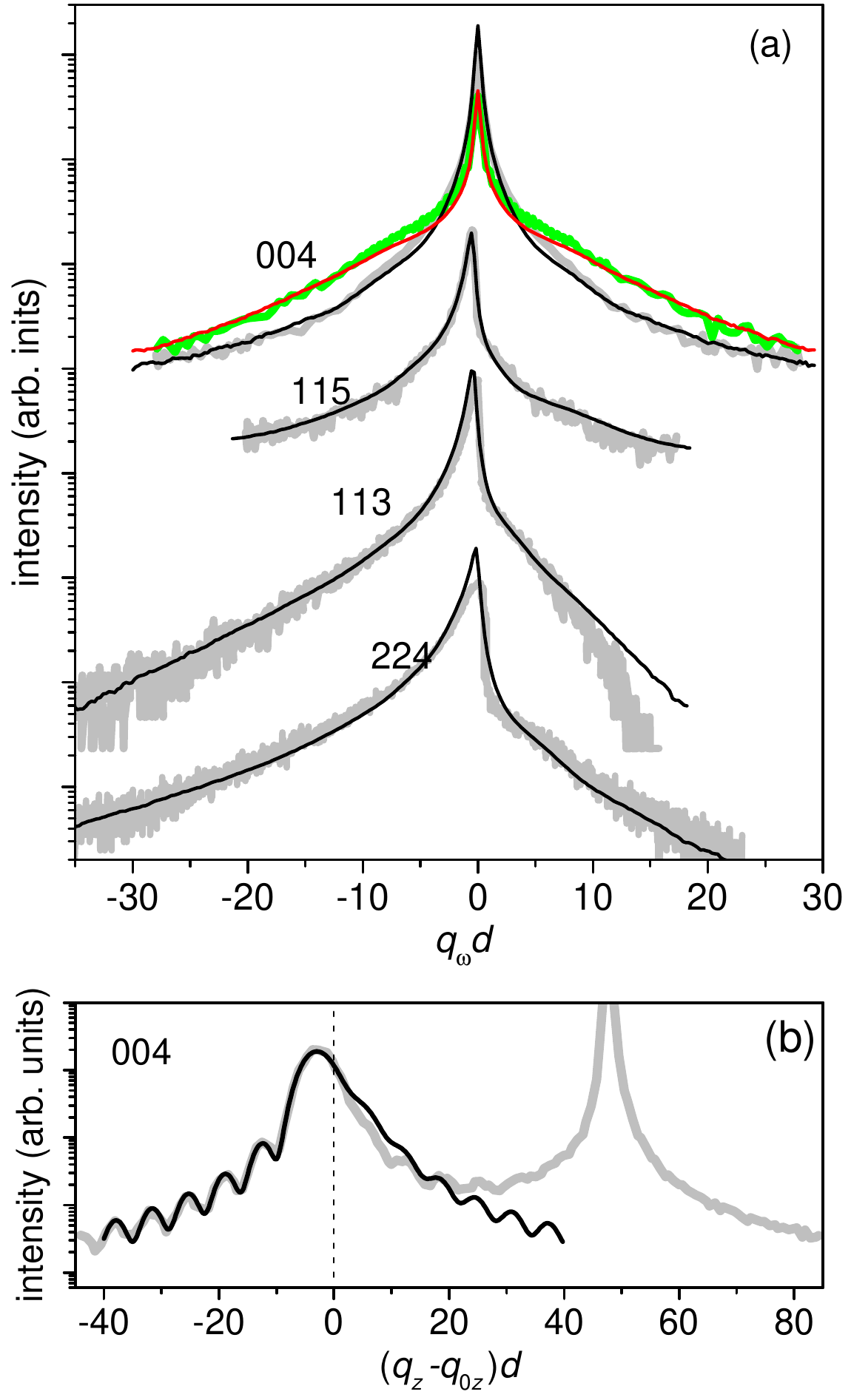}
\caption{Scans (a) perpendicular to the diffraction vectors ($\omega$-scans)
and (b) along the diffraction vector 004 ($\omega-2\theta$ scan).
The experimental scans are shown by thick gray lines and the Monte
Carlo calculated ones by black lines. For 004 reflection, two $q_{x}$-scans
are plotted. The scan at the position of the coherent maximum on the
map in Fig.~\ref{fig:maps}(a) is shown by the gray line, while the
scan at the maximum of diffuse scattering by the green line.}
\label{fig:scans}
\end{figure}

Figure \ref{fig:scans}(a) shows line scans extracted
from the maps perpendicular to the scattering vectors
($\omega$-scans) at the intensity maxima of the respective maps.
For the 004 reflection, two scans are presented, one through the maximum
of the coherent intensity (gray line) and the other through the maximum
of the diffuse intensity (green line). Evidently, the former scan
shows a larger peak intensity, while the latter has a higher diffuse
scattering intensity. In asymmetric reflections, the scans through the coherent
and the diffuse maxima reveal a less pronounced (albeit present) difference,
and we present only the scans through the coherent maxima.

The reciprocal space maps in asymmetric reflections in Figs.~\ref{fig:maps}(b-d)
reveal a strong asymmetry of the diffuse intensity distributions.
For each reflection, the coherent streak at $q_{x}=0$ separates the diffuse
intensity in two lobes, the one at $q_{x}<0$ possessing notably higher
intensity in comparison with the other at $q_{x}>0$. This asymmetry
and a sharp border between the two lobes is clearly seen in the scans
presented in Fig.~\ref{fig:scans}(a).

Our aim now is to find a distribution of misfit dislocations that
complies with all characteristic features of the experimental diffraction
patterns in Figs\@.~\ref{fig:maps} and \ref{fig:scans}. The 60$^\circ$
dislocations with Burgers vectors $\nicefrac{1}{2}\left\langle 011\right\rangle $
that glide in the $\{111\}$ planes are the main type of dislocations
in crystals with diamond or zinc blende structure 
\cite{fitzgerald91,hull92,jain97,bolkhovityanov01}.
Calculated and experimental diffraction curves from uncorrelated 60$^\circ$
dislocations show characteristic side peaks (side lobes on the maps)
\cite{kidd95,kaganer97,alexe02,alexe05} that are absent in our experimental
patterns. Since edge (Lomer type) misfit dislocations are also formed
at the SiGe/Si interface as a result of a reaction between 60$^\circ$
dislocations \cite{bolkhovityanov12,bolkhovityanov13,marzegalli13},
we have tried various models for the arrangement of edge dislocations,
using the Monte Carlo method \cite{kaganer09prbMC}. However, the
calculated profiles are notably narrower than our experimental profiles.

A Monte Carlo method for the calculation of the x-ray diffraction
intensity from relaxed epitaxial films has been formulated in 
Ref.~\cite{kaganer09prbMC}. The method is applicable to any kind of 
misfit dislocations but used so far only
for edge (Lomer type) dislocations in symmetric Bragg reflections.
The positional correlations of dislocations were considered implying
that the dislocations tend from random to more regular arrangements
to reduce the elastic energy of the film. In the present work, we have
included 60$^\circ$ dislocations and asymmetric reflections,
and searched in a wider range of possible positional correlations.

Reciprocal space maps calculated by the Monte Carlo method 
in Figs.~\ref{fig:maps}(e-h)
and the diffraction profiles shown by black lines in Fig.~\ref{fig:scans}
demonstrate a quantitative agreement with the experimental maps and
curves. Now we describe the Monte Carlo model of the dislocation distribution
that we have used. We assume two arrays of straight misfit dislocations
with the dislocation lines in the two orthogonal $\left\langle 110\right\rangle $
directions. For dislocations with the lines normal to the scattering
plane $(x,z)$, Burgers vectors $\mathbf{b}=\nicefrac{1}{2}\left\langle 011\right\rangle $
have the same component $b_{x}=-a/2\sqrt{2}$ releasing the misfit,
while the signs of two other components, screw $b_{y}=\pm a/2\sqrt{2}$
and edge $b_{z}=\pm a/2$, are chosen on random and uncorrelated (here
$a$ is the lattice parameter of the substrate). The position of the
Bragg peak corresponding to the average relaxation is given by $q_{0x}=\rho Q_{x}b_{x}$
and $q_{0z}=-\frac{2\nu}{1-\nu}\rho Q_{z}b_{x}$, where $\rho$ is
the linear density of misfit dislocations and $Q_{x},Q_{z}$ are the
components of the reciprocal lattice vector \cite{kaganer97}. The
shift $q_{0z}$ is taken into account in Figs.~\ref{fig:maps}(e-h),
but the $q_{0x}$-shift is not made, as discussed below.

To reach an agreement between the experimental and the calculated
curves in Figs.~\ref{fig:maps} and \ref{fig:scans}, we varied the
dislocation density and the distribution of dislocations. The dislocation
density $\rho d=0.5$ used in the Monte Carlo calculations corresponds
to a relaxation degree $R=0.05$. The positions of the dislocations
are modeled as a Markov chain, with the probability $P$ to have a
distance $\rho^{-1}P$ between two subsequent dislocations possessing
a lognormal distribution. The probability density is generated as
$P=\exp[\mu+\sigma N(0,1)]$, where $N(0,1)$ is the standard normal
distribution with the mean 0 and dispersion 1. The standard deviation
of the lognormal distribution is taken to be $s=10$ times larger
than its mean value. Explicitly, the parameters of the lognormal distribution
are $\sigma=\sqrt{\ln(1+s^{2})}$=2.15 and $\mu=-\sigma^{2}/2=-2.31$.

\begin{figure}
\includegraphics[width=1\columnwidth]{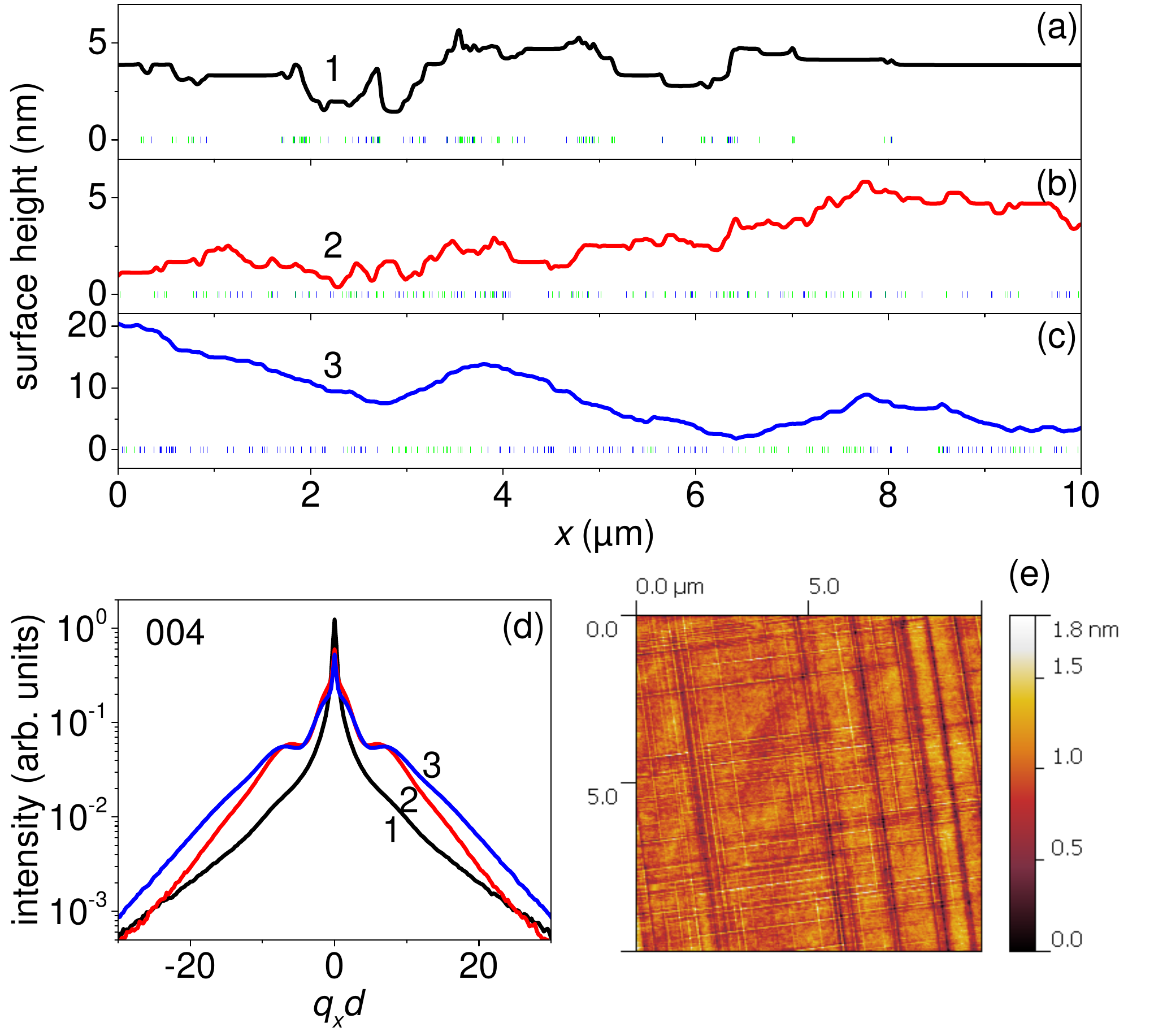}
\caption{(a-c) Examples of dislocation positions and surface profiles for different
models of the dislocation arrangements. Blue and green bars represent
dislocations with two different tilt components $\pm b_{z}$ of the
Burgers vectors of 60$^\circ$ dislocations. (a) The
model reproducing the experimental maps and profiles in Figs\@.~\ref{fig:maps}
and \ref{fig:scans}: lognormal distribution of the dislocation positions
with the standard deviation 10 times larger than the average distance
between dislocations, random uncorrelated Burgers vectors. (b) Both
positions and Burgers vectors are uncorrelated. (c) Uncorrelated dislocation
positions and groups of dislocations with the same Burgers vectors,
10 dislocations in a group on average. (d) Diffraction profiles in
symmetric reflection 004 for the three models above. (e) Atomic force
microscopy image of the experimental sample showing the cross-hatch
pattern. }
\label{fig:surface}
\end{figure}

Figure \ref{fig:surface}(a) shows an example of the dislocations
distributed according to our model, and the surface displacement caused
by these dislocations. The positions of the dislocations are marked by
vertical bars. Blue and green bars correspond to dislocations with
opposite signs of the tilt component of the Burgers vector $b_{z}$.
The large standard deviation of the distribution gives rise to bunches
of dislocations separated by dislocation-free regions. The black curve
in Fig\@.~\ref{fig:surface}(a) is the surface displacement calculated
for this dislocation array. On a mesoscopic scale, the surface relief
exhibits rather sharp peaks caused by dislocation bundles separated
by relatively flat regions. In calculating surface displacements from
dislocations, we assume that the slip steps are eliminated by surface
diffusion, as proposed by Andrews \emph{et al.} \cite{andrews00,andrews02,andrews04},
so that the surface displacement due to each dislocation is a continuous
function of the coordinate $x$. The x-ray diffraction profile, calculated
for this distribution of the dislocation positions, is shown in Fig.~\ref{fig:surface}(d)
by the black line. It is calculated at $q_{z}=q_{0z}$, i.e., at the $q_{z}$
position in between the ones presented in Fig.~\ref{fig:scans}(a),
for the same model of the dislocation distribution.

Figures \ref{fig:surface}(b,c) present, for a comparison, more homogeneous
dislocation distributions modeled in the literature \cite{andrews00,andrews02,andrews04}.
The dislocation positions in Fig.~\ref{fig:surface}(b) are chosen
on random independently from each other, and the signs of the tilt
components of their Burgers vectors are also not correlated. The dislocation
density is the same, $\rho d=0.5$. The surface displacements (red
line) are qualitatively similar to those in Fig.~\ref{fig:surface}(a)
but have less pronounced peaks and smaller flat areas. However, the
x-ray diffraction profile calculated for this model by the same Monte
Carlo method {[}red curve in Fig.~\ref{fig:surface}(d){]} looks
qualitatively different. It possesses the side maxima described theoretically
for uncorrelated random misfit dislocations \cite{kaganer97} and
observed experimentally for In$_{0.1}$Ga$_{0.9}$As/GaAs(001) \cite{kidd95},
Si$_{0.75}$Ge$_{0.25}$/Si(001) \cite{kaganer97}, and ZnSe/GaAs(001)
\cite{alexe02,alexe05} epitaxial films. 

Figure \ref{fig:surface}(c) shows a model with random dislocation
positions but correlated Burgers vectors. Alternating groups containing
a given number of dislocations with the same tilt component of the
Burgers vector $b_{z}$ were considered in Ref.~\cite{andrews04}.
In our model, the number of dislocations in a group is taken on random:
the sign of $b_{z}$ is changed with the probability $p=0.1$, so
that there are on average 10 dislocations in a group. The surface
profile (blue line) varies over a larger lateral scale and a larger
height. The diffraction profile calculated
for this model {[}blue line in Fig.~\ref{fig:surface}(e){]} exhibits
the same side maxima as the one for the case of uncorrelated positions
and Burgers vectors above. Thus, the bunching of dislocations is required
to explain our experimental diffraction profiles.

Figure \ref{fig:surface}(e) presents the surface relief of the investigated
Si$_{0.4}$Ge$_{0.6}$/Si(001) film as measured by atomic force microscopy
(AFM). The cross-hatch pattern observed in this micrograph is 
a well-known manifestation of plastic relaxation \cite{olsen75,lutz95,gallas99,andrews00,andrews02,andrews04,saha13}.
With the dislocation density $\rho d=0.5$ as determined from the
x-ray data, the number of dislocations on a 10~\textmu m interval
is 185, while only about 35 randomly spaced parallel lines are seen
in Fig.~\ref{fig:surface}(e). Hence, a single line in the AFM image
corresponds to a group of dislocations, rather than a single dislocation.
This is in a good agreement with our analysis and the calculated profile
in Fig.~\ref{fig:surface}(a). The formation of the cross-hatch pattern
is a result of a complicated interplay between surface diffusion and
dislocation generation. It remains debatable, if the dislocations
cause the surface undulations or, oppositely, the undulations due
to surface diffusion are sources of dislocations \cite{saha13}. Hence,
we do not make a quantitative comparison of the measured and the modeled
surface profiles.

The expected shift $q_{0z}$ of the coherent 004 peak due to an average
strain calculated by the expression given above for the dislocation
density of our sample $(\rho d=0.5)$ is equal to $q_{0z}d\approx4.4$.
This shift is taken into account in Fig.~\ref{fig:scans}(b), so
that the intensity maximum is expected to be at the origin, $q_{z}-q_{0z}=0$.
We have verified this prediction by additional Monte Carlo calculations
(not shown) of similar diffraction profiles for uncorrelated or
more ordered dislocations, which give the intensity maxima at the
expected position. However, the peak of the calculated curve in Fig.~\ref{fig:scans}(b)
is at $(q_{z}-q_{0z})d\approx-2.65$. Hence, the coherent peak position
corresponds to less than half of the actual film relaxation.

The difference between expected and calculated peak positions can
be explained by the dislocation bunching, which gives rise to regions
with large and small strains, as it is reflected in the surface profile
in Fig.~\ref{fig:surface}(a). The dislocation-rich regions possess
large strain and large strain inhomogeneity, so that they contribute
mostly to the diffuse scattering. In contrast, the dislocation-depleted
regions possessing small strain and small strain gradients contribute
to the coherent intensity. As a result, the positions of the coherent
and the diffuse intensity maxima on the calculated reciprocal space
maps in Figs.~\ref{fig:maps}(e-h) do not coincide: the coherent
peak reflects the areas in the sample which are less strained than
the average, while the diffuse peak represents the more strained ones.

The coherent peaks in Figs.~\ref{fig:maps}(f-h) remain at the same
lateral position $q_{x}=0$ as they are in an elastically relaxed
dislocation-free film. This peak position has been analyzed in Ref.~\cite{kaganer97}
{[}see discussion after Eq.~(27){]} and in Ref.~\cite{benediktovitch11pss}.
One can also see from the experimental maps in the insets in Figs.~\ref{fig:maps}(a-d),
that the $q_{x}$-positions of the substrate and the film peaks coincide.
The intensity maxima move to $q_{x}=q_{0x}$ when the dislocation
density is increased, the coherent peak weakens, and the diffuse peak
dominates.

Summarizing, diffuse x-ray intensity from misfit dislocations can
be revealed at the very early stages of relaxation of the epitaxial
films, when the shift of diffraction peaks due to these dislocations
is not yet visible. The diffuse intensity distribution is sensitive
to the spatial arrangement of misfit dislocations. We model the dislocation
distribution by the Monte Carlo method and find that the diffraction
pattern from a Si$_{0.4}$Ge$_{0.6}$/Si(001) epitaxial film on the
onset of relaxation is due to a very inhomogeneous dislocation distribution.
Distances between dislocations vary very broadly, so that the standard
deviation of the dislocation spacings is 10 times larger than
the mean distance between dislocations. In other words, dislocations
form bunches, as a result of the action of small number of dislocation
sources. These bunches are seen as cross-hatch patterns in the AFM
images of the film.

The inhomogeneous dislocation distribution results in peculiar features
of the diffraction patterns. The positions of the coherent and the
diffuse peaks do not coincide, since the former is mostly due to undisturbed
regions between dislocation bunches while the latter is due to the
inhomogeneous strain at the bunches. Moreover, since the coherent
peak represents the undisturbed regions, rather than the strain averaged
over the whole film, its position at the onset of relaxation does
not correspond to the actual density of misfit dislocations and underestimates
relaxation by more than a factor of 2.

The authors thank Bernd Jenichen and Oliver Brandt for a critical reading of 
the manuscript.


%

\end{document}